\documentclass[a4paper,11pt]{article}
\usepackage{jheppub}
\usepackage[latin1]{inputenc} 
\usepackage[T1]{fontenc}
\usepackage[safe]{textcomp}
\usepackage{lmodern} 
\usepackage{epsfig}
\usepackage{float}
\usepackage{amstext}  
\usepackage{amsfonts}
\usepackage{amsthm}
\usepackage{bm}       
\usepackage[bottom]{footmisc} 
\usepackage{array}                
\usepackage{xcolor} 
\usepackage{framed}
\usepackage{slashed}
\usepackage[absolute]{textpos}
\usepackage{axodraw4j}
\usepackage{dsfont}
\usepackage{multirow}

\newcommand{\p}{\partial}
\newcommand{\ddp}[1]{\frac{\partial}{\partial#1}}
\newcommand{\f}[2]{\frac{#1}{#2}}
\newcommand{\sss}[1]{\scriptscriptstyle#1}
\newcommand{\ssst}[1]{\scriptscriptstyle{\text{#1}}}

\newcommand{\bea}{\begin{eqnarray}}
\newcommand{\eea}{\end{eqnarray}}
\newcommand{\be}{\begin{equation}}
\newcommand{\ee}{\end{equation}}
\newcommand{\ba}{\begin{align}}
\newcommand{\ea}{\end{align}}
\newcommand{\beas}{\begin{eqnarray*}}
\newcommand{\eeas}{\end{eqnarray*}}
\newcommand{\bes}{\begin{equation*}}
\newcommand{\ees}{\end{equation*}}
\newcommand{\bas}{\begin{align*}}
\newcommand{\eas}{\end{align*}}

\newcommand{\ssL}{{\mathcal L}} 
\newcommand{\eps}{{\varepsilon}}
\newcommand{\cd}{{\cdot}} 
\newcommand{\cf}{C_{\scriptscriptstyle{F}}} 
\newcommand{\ca}{C_{\scriptscriptstyle{A}}}
\newcommand{\tr}{T_{\scriptscriptstyle{F}}}

\newcommand{\Ng}{n_{\scriptscriptstyle{g}}}

\newcommand{\Nf}{n_{\scriptscriptstyle{f}}}
\newcommand{\gs}{g_{\scriptscriptstyle{s}}}
\newcommand{\gsB}{g_{\scriptscriptstyle{s}}^{\scriptscriptstyle{\text{B}}}}

\newcommand{\als}{\alpha_{\scriptscriptstyle{s}}}
\newcommand{\as}{a_{\scriptscriptstyle{s}}}
\newcommand{\lb}{\left(}
\newcommand{\rb}{\right)}

\definecolor{bluemar}{rgb}{0,0,.5}
\definecolor{redmar}{rgb}{.8,0,0}
\definecolor{greenmar}{rgb}{0,.5,0}

\def\bbuildrel#1_#2^#3%
{\mathrel{\mathop{\kern 0pt#1}\limits_{#2}^{#3}}}

\newcommand{\ice}[1]{\relax}
\newcommand{\beq}{\begin{equation}}
\newcommand{\eeq}{\end{equation}}

\title{On the renormalization of operator products: the scalar gluonic case}
\author[a]{Max F. Zoller}
\affiliation[a]{Institut f\"ur Physik, University of Zurich (UZH), Winterthurerstrasse 190,
CH-8057 Zurich, Switzerland}

\emailAdd{zoller@physik.uzh.ch}

\abstract{
In this paper we study the renormalization of the product of two operators 
\mbox{$O_1=-\f{1}{4} G^{\mu \nu}G_{\mu \nu}$} in QCD.
An insertion of two such operators $O_1(x)O_1(0)$ into a Greens
function produces divergent contact terms for $x\rightarrow 0$.

In the course of the computation of the operator product expansion (OPE) of the
correlator of two such operators \mbox{$i\int\!\mathrm{d}^4x\,e^{iqx} T\{\,O_1(x)O_1(0)\}$} to three-loop 
order \cite{Zoller:2012qv,Zoller:2014dca} we discovered that divergent contact terms remain
not only in the leading Wilson coefficient $C_0$, which is just the VEV of the correlator, but
also in the Wilson coefficient $C_1$ in front of $O_1$. As this correlator plays an important role for 
example in QCD sum rules a full understanding of its renormalization is desireable.

This work explains how the divergences encountered in higher orders of an OPE of this correlator
should be absorbed in counterterms and derives an additive renormalization constant for $C_1$ from
first principles and to all orders in perturnbation theory. 
The method to derive the renormalization of this operator product is an extension of the 
ideas of \cite{Spiridonov:1984br} and can be generalized to other cases.
}

\arxivnumber{}
\subheader{ZU-TH-4/16}

\begin{document}
\maketitle

\setlength{\fboxrule}{0.5 mm} 

\section{Introduction: The scalar gluonic operator $O_1$ and its correlator}

Local operators, i.e. products of fields at the same point in space-time, play an important role in quantum field
theory (QFT) as they serve as building blocks for Lagrangians and Greens functions. 
The bilocal correlator of two such local operators is an important object in applications of QFT, such as sum rules.
In this paper we study the renormalization of the scalar gluonic operator 
\be 
O_1(x) :=-\f{1}{4} G^{a\,\mu \nu}G^{a}_{\mu \nu}(x) \label{O1def}
\ee
constructed from the field strength tensor of QCD
\be
G_{\mu \nu}^{a}=
\p_\mu A_\nu^{a} - \p_\nu A_\mu^{a} 
+ \gs f^{abc} A_\mu^{b} A_\nu^{c}
{}.
\ee
The operator \eqref{O1def} appears in the massless QCD Lagrangian
\be
\begin{split}
\ssL=&-\f{1}{4}\,G^a_{\mu \nu} G^{a\,\mu \nu}-\f{1}{2\lambda}\lb\p_\mu A^{a\,\mu}\rb^2 
+\p_\rho \bar{c}^a \p^{\rho}c^a+\gs f^{abc}\p_\rho \bar{c}^a A^b c^c \\
& +\bar{\psi}\lb\f{i}{2}\overleftrightarrow{\slashed{\p}}-m\rb\psi + \gs\bar{\psi}\slashed{A}^a T^a\psi
\end{split}
\label{QCDLag} 
\ee
where $T^a$ are the generators and $f^{abc}$ the structure constants of the gauge group.

A renormalized version of this operator, i.e. one which gives finite results if inserted into a Greens function,
was obtained in \cite{Nielsen:1975ph,Spiridonov:1984br}. If we only consider matrix elements with physical external states
and $m=0$ it can be renormalized multiplicatively:
\be
[O_1]=Z_{11} O_1^{\ssst{B}}=-\f{Z_{11}}{4}G^{{\ssst{B}}\,a\,\mu \nu}G^{\ssst{B}\,a}_{\mu \nu}{} \label{O1ren}
\ee
where $[\ldots]$ marks the renormalized operator and the index B bare quantities, which means that all fields and couplings are
replaced by bare ones.
The renormalization constant derived in \cite{Nielsen:1975ph,Spiridonov:1984br}
\be Z_{11}=1+\als\f{\p}{\p\als}\ln Z_{\als}=\lb 1- \f{\beta(\als)}{\eps}\rb^{-1} \label{Z11fromZa}\ee
can be expressed through the beta function
\beq\beta(\als)=
\mu^2\f{\mathrm{d}}{\mathrm{d}\mu^2}\, \ln \als
=  - \sum_{i \ge 0} \beta_i \, \left( \frac{\als}{\pi} \right)^{i+1}
{}\label{be:def}
\eeq
and at first order in $\als$ is equal to the renormalization constant for $\als$: $Z_{11}=Z_{\als}+\mathcal{O}(\als^2)$.

The bilocal correlator of this operator is defined as
\be 
\widehat{\Pi}^{\ssst{GG}}(q^2):=i\int\!\mathrm{d}^4x\,e^{iqx} T\{\,[O_1](x)[O_1](0)\}=
i\,Z_{11}^2\int\!\mathrm{d}^4x\,e^{iqx} T\{\,O_1^{\ssst{B}}(x)O_1^{\ssst{B}}(0)\}{},
\label{defPihat}
\ee

The OPE of the correlator \eqref{defPihat} (considering only scalar operators) reads
\bea
\widehat{\Pi}^{\ssst{GG}}(q^2) &=&  
q^4\,  C_0^{\ssst{GG}}(q^2)\, 1 
\ \ + \ \  C_1^{\ssst{GG}\,\ssst{B}}(q^2) \, O_1^{\ssst{B}} \ \ + \ \ 
\sum\limits_{i} C_i^{\ssst{GG}\,\ssst{B}}(q^2) O_i^{\ssst{B}}\ \ + \ \ 
\mathcal{O}\lb\f{1}{q^2}\rb  \label{O1O1OPE} \\
&=& q^4\,  C_0^{\ssst{GG}}(q^2)\, 1 
\ \ + \ \  C_1^{\ssst{GG}}(q^2) \, [O_1] \ \ + \ \ 
\sum\limits_{i} C_i^{\ssst{GG}}(q^2) [O_i]\ \ + \ \ 
\mathcal{O}\lb\f{1}{q^2}\rb, \nonumber
\eea
where the sum goes over a set of mass dimension four operators which form a suitable basis 
together with $O_1^{\ssst{B}}$ or $[O_1]$ (see section \ref{Sec:renO1}).
For sum rules (see e.g.~\cite{forkel_sumrule}) we are usually interested in the vacuum expectation value (VEV) of the correlator
\be 
 q^4\,\Pi^{\ssst{GG}}(q^2)=\langle 0|\widehat{\Pi}^{\ssst{GG}}(q^2)|0 \rangle 
\label{defPiVEV}
\ee
for large Euclidean momenta $-q^2\gg 0$. As the VEV of unphysical operators\footnote{These are gauge dependent operators or operators
which vanish due to equations of motion.} vanishes we can restrict ourselves to
physical operators.

$C_0^{\ssst{GG}}$ is known at four-loop level from \cite{Baikov:2006ch} and $C_1^{\ssst{GG}}$ at three-loop level from \cite{Zoller:2014dca}.
In \cite{Zoller:2012qv,Zoller:2014dca} it was discovered, however, that the described renormalization procedure does not
yield a finite result for $C_1^{\ssst{GG}}$ starting from two-loop level. These divergent terms are proportional
to $\delta^{(4)}(x)$ in x-space and hence stem from the point where both operators $O_1$ in the correlator \eqref{defPihat}
are at the same point $x=0$. For this reason they are called contact terms.\footnote{In the case of the correlator 
of the pseudoscalar operator $\tilde{O}_1(x)  :=G^{a\,\mu \nu}G^{a\,\rho \sigma}\eps_{\mu \nu\rho\sigma}$ it was 
proven in \cite{Zoller:2013ixa} that no contact terms can apear in $C_1$ which was explicitly confirmed in a 
three-loop calculation of this quantitiy in the same paper.}

The complete renormalization of the operator product of two $O_1$ is also desirable for phenomenological applications in effective theories.
An important example is double Higgs production in the framework of an effective theory with $m_t\to\infty$. Having integrated out the top loops
the resulting vertices are $\propto O_1 H$ and $\propto O_1 HH$, where $H$ is the Higgs field (see e.g. \cite{Grigo:2014jma}). 
the new renormalization constant $Z_{11}^{\ssst{L}}$ defined below in \eqref{C1finitecorrO1O1} will be needed in a counterterm $\propto O_1 HH$
if two effective vertices $\propto O_1 H$ are inserted into two-loop diagrams with two external gluons, i.e. starting at one loop-order
higher than the results presented in \cite{Grigo:2014jma}.

%
The paper is structured as follows: In section \ref{Sec:renO1} the renormalization of Greens functions with one insertion of $O_1$ is reviewed
following the ideas of \cite{Spiridonov:1984br}. This method is then extended in section \ref{Sec:renO1O1} in order to renormalize the product of two such operators
followed by the application of the found result to the OPE \eqref{O1O1OPE} in section \ref{Sec:renO1O1} explaining the contact term in $C_1^{\ssst{GG}}$ 
\cite{Zoller:2012qv,Zoller:2014dca}. We finish with some conclusions and acknowledgments.

\section{Renormalization of $O_1$} \label{Sec:renO1}

For the purpose of this and the next section we rescale the field \mbox{$A_\mu^a\rightarrow \f{A_\mu^a}{g}$} transforming \eqref{QCDLag} into
\be
\begin{split}
\ssL=&-\f{1}{4\gs^2}\,G^a_{\mu \nu} G^{a\,\mu \nu}-\f{1}{2\lambda\gs^2}\lb\p_\mu A^{a\,\mu}\rb^2 
+\p_\rho \bar{c}^a \p^{\rho}c^a+f^{abc}\p_\rho \bar{c}^a A^b c^c \\
&+\bar{\psi}\lb\f{i}{2}\overleftrightarrow{\slashed{\p}}-m\rb\psi + \bar{\psi}\slashed{A}^a T^a\psi
\end{split}
\label{QCDLagresc} 
\ee
with the rescaled field strength tensor
\be
G_{\mu \nu}^{a}=
\p_\mu A_\nu^{a} - \p_\nu A_\mu^{a} 
+ f^{abc} A_\mu^{b} A_\nu^{c}
{}.
\ee
We define the renormalization prescriptions
\be A^{\ssst{B}\,a}_\mu=A^{a}_\mu\f{Z_1}{Z_3}, \quad c^{\ssst{B}\,a}=c^{a}\sqrt{\tilde{Z}_3}, \quad 
\psi^{\ssst{B}}=\psi \sqrt{Z_2}, \quad m^{\ssst{B}}=m Z_m, \quad
\gs^{\ssst{B}}=\gs \f{Z_1}{Z_3^{3/2}}\equiv \gs Z_g, \ee
where $Z_g^2=Z_{\als}$ with $\als=\f{\gs^2}{4\pi}$.
Hence \be \f{A^{\ssst{B}\,a}_\mu}{\gs^{\ssst{B}}}=\f{A^{a}_\mu}{\gs}\sqrt{Z_3} \ee
which is just the renormalization procedure for $A_\mu^a$ in the original Lagrangian \eqref{QCDLag}. 
$Z_1$, $Z_2$, $Z_3$, $\tilde{Z}_3$ and $Z_m$ are therefore the usual renormalization constants of QCD.\footnote{ 
The same as with unrescaled $A_\mu^a$ and the definition $A^{\ssst{B}\,a}_\mu=A^{a}_\mu\sqrt{Z_3}$.}
The bare Lagrangian reads
\be
\begin{split}
\ssL_{\ssst{B}}=&-\f{1}{4(\gsB)^2}\,G^{\ssst{B}\,a}_{\mu \nu} G^{\ssst{B}\,a\,\mu \nu}-\f{1}{2\lambda\,Z_3\,(\gsB)^2}\lb\p_\mu A^{\ssst{B}\,a\,\mu}\rb^2 
+\p_\rho \bar{c}^{\ssst{B}\,a} \p^{\rho}c^{\ssst{B}\,a}\\ &+f^{abc}\p_\rho \bar{c}^{\ssst{B}\,a} A^{\ssst{B}\,b} c^{\ssst{B}\,c}  
+\bar{\psi}^{\ssst{B}}\lb\f{i}{2}\overleftrightarrow{\slashed{\p}}-m_{\ssst{B}}\rb\psi^{\ssst{B}} + \bar{\psi}^{\ssst{B}}\slashed{A}^{a\,\ssst{B}} T^a\psi^{\ssst{B}}
{}.
\end{split}
\label{QCDLagrescB} 
\ee
Finite results for Greens functions are usually obtained by applying the R-Operation (see e.~g.~\cite{Kennedy_ROp,collins}) to
the unrenormalized Greens function or equivalently by using the bare Lagrangian in which a counterterm for every
operator in the Lagrangian is defined. Finite Greens functions are derived from the generating functional of the path integral formalism
\begin{eqnarray} 
Z^{\ssst{R}} =& \textbf{R}\,& \int\!\mathrm{d}\Phi \,\mathrm{e}^{i\int\!\mathrm{d}^4x\,(\ssL+J\cdot\Phi)}   \label{ZR1}\\
=& & \int\!\mathrm{d}\Phi \,\mathrm{e}^{i\int\!\mathrm{d}^4x\,(\ssL_{\ssst{B}}+J\cdot\Phi)}  \label{ZR2} 
\end{eqnarray}
with the multiplets of all fields in the Lagrangian and the respective external currents
\be \Phi:=(\f{A^\mu}{\gs},\bar{c},c,\bar{\psi},\psi),\quad J:=(J_\mu,J_{\bar{c}},J_{c},J_{\bar{\psi}},J_{\psi}) \ee 
and the integration measure
\be
\mathrm{d}\Phi:=\mathrm{d}A^\mu \,\mathrm{d}\bar{c} \,\mathrm{d}c\, \mathrm{d}\bar{\psi} \,\mathrm{d}\psi{}.
\ee
A finite Greens function with the insertion of a local operator \mbox{$\widetilde{O}_i(q)=\int\!\mathrm{d}^4x\, e^{iqx}O_i(x)$} is obtained as
\be
Z_{O_i}^R(q)=\textbf{R}\,\int\!\mathrm{d}\Phi \,\widetilde{O}_i(q)\, e^{i\int\!\mathrm{d}^4x\lb \ssL+J\Phi \rb}
\label{ZOiR1}
\ee
which can also be written in terms of the bare Lagrangian and a superposition of bare local operators
\be
Z_{O_i}^R=Z_{ij}\int\!\mathrm{d}\Phi \,\widetilde{O}_j^B(q)\, e^{i\int\!\mathrm{d}^4x\lb \ssL_{\sss{B}}+J\Phi \rb}
\label{ZOiR2}
\ee
In MS-like schemes the renormalization constants for these operators do not depend on $q$ and hence we set $q=0$.
The important point now is that in renormalization schemes based on minimal subtraction the R-Operation commutes
with the operation of taking derivatives wrt the parameters of the theory $\gs,\lambda,\ldots$ and wrt external 
currents. An operator insertion of \mbox{$\widetilde{O}_1\equiv\widetilde{O}_1(0)$} in a Greens function can be obtained \cite{Spiridonov:1984br} 
by applying the operation 
\be \textbf{D}_1 :=\f{1}{i}\lb \lambda\ddp{\lambda}-\f{1}{2}\gs\ddp{\gs}-\f{1}{2}J_\mu\f{\delta}{\delta J_\mu}\rb
\ee
to \eqref{ZR1}:
\bea \textbf{D}_1\,Z^{\ssst{R}}&=& \textbf{D}_1\, \textbf{R}\, 
\int\!\mathrm{d}\Phi \,\mathrm{e}^{i\int\!\mathrm{d}^4x\,(\ssL+J\cdot\Phi)} =
\textbf{R}\, \textbf{D}_1\,
\int\!\mathrm{d}\Phi \,\mathrm{e}^{i\int\!\mathrm{d}^4x\,(\ssL+J\cdot\Phi)}\label{D1ZR1}\\
&=&\textbf{R}\, \int\!\mathrm{d}\Phi \,\widetilde{O}_1\,\mathrm{e}^{i\int\!\mathrm{d}^4x\,(\ssL+J\cdot\Phi)}\nonumber
 \eea
Using the representation \eqref{ZR2} of $Z_{\ssst{R}}$ we find that this equals
\bea 
\textbf{D}_1\,Z^{\ssst{R}}&=&\int\!\mathrm{d}\Phi \,
\left\{\lb \lambda\ddp{\lambda}-\f{1}{2}\gs\ddp{\gs}\rb\ssL_{\ssst{B}}\right\}
\mathrm{e}^{i\int\!\mathrm{d}^4x\,(\ssL_{\ssst{B}}+J\cdot\Phi)} \label{D1toLB}\\
&=&\int\!\mathrm{d}\Phi \,\left\{\sum\limits_{i=1}^{5} Z_{1i}\widetilde{O}_i^{\ssst{B}}\right\}\,
\mathrm{e}^{i\int\!\mathrm{d}^4x\,(\ssL_{\ssst{B}}+J\cdot\Phi)}\label{D1toLBZi}
\eea
A suitable basis of mass dimension four operators was given in \cite{Spiridonov:1984br}:
\bea
O_1 &=& -\f{1}{4\gs^2}G^{a\,\mu \nu}G^{a}_{\mu \nu} \label{O1Sp},\\
O_2 &=& m\bar{\psi}\psi \label{O2Sp},\\
O_3 &=& \bar{\psi}\lb \f{i}{2}\overleftrightarrow{\slashed{\p}}-m \rb\psi \label{O3Sp},\\
O_4 &=& A^a_\nu \left\{ \lb \delta^{ab}\p_\mu-gf^{abc}A^c_\mu \rb G^{b\,\mu\nu} 
         +\bar{\psi}T^a \gamma^\nu \psi \right\} \label{O4Sp}\\
         & & -\lb\p_\mu\bar{c^a}\rb\lb\p^\mu c^a\rb, \nonumber \\
O_5 &=& \left\{\lb \delta^{ab}\p_\mu-f^{abc}A^c_\mu \rb\p^\mu\bar{c^a}\right\}c^b \label{O5Sp}.
\eea
Using \eqref{D1toLB} and collecting the coefficients in \eqref{D1toLBZi} we find the 
renormalization constants $Z_{1i}$. The second line of \eqref{D1ZR1} gives us the renormalized operator $[O_1]$, such that
from \eqref{D1ZR1} and \eqref{D1toLBZi} we have $[O_1]=Z_{1j}O_j^{\ssst{B}}$.
Similarly the renormalization constants for the other
operators \be [O_i]=Z_{ij}O_j^{\ssst{B}}. \ee are derived:
\be Z_{ij}=\delta_{ij}+\overline{D}_i\ln \overline{Z}_j \qquad (i,j\in \{1,\ldots,5\})  \label{Zijformel} \ee
with
\bea
\overline{D}_1 &=& \lambda\f{\p}{\p\lambda}-\als \f{\p}{\p\als},\;\;  \overline{D}_2 =-m\f{\p}{\p m}, \;\;
\overline{D}_3 = 0, \;\; \overline{D}_4 =2\lambda \f{\p}{\p\lambda}, \;\; \overline{D}_5=0, \\
\overline{Z}_1 &=& Z_\alpha^{-1}, \;\;\overline{Z}_2=Z_m^{-1}, \;\;\overline{Z}_3=Z_2, \;\;\overline{Z}_4 =Z_1 Z_3^{-1}, \;\;
\overline{Z}_5 =Z_3 Z_1^{-1}. \eea
These were first found in \cite{Spiridonov:1984br} and rederived for this study.\footnote{For the coefficients 
of the unphysical operators an additional ``counting identity'' is needed for which we refer to \cite{Spiridonov:1984br}.}
The gauge-invariant operators \eqref{O1Sp} and \eqref{O2Sp} are physical operators of class $I$ according to the classification
from \cite{Deans:1978wn,Spiridonov:1984br}. In physical matrix elements the class $I$ operators do not vanish whereas
the gauge-invariant class $II^a$ operator \eqref{O3Sp} vanishes due to an equation of motion. The non gauge-invariant operators
\eqref{O4Sp} and \eqref{O5Sp} are of class $II^b$ and vanish due to a BRST identity in physical matrix elements.
Hence in the massless case $O_1$ is renormalized multiplicatively with $Z_{11}$ as given in  
\eqref{Z11fromZa}. In the following we set $m=0$.

\section{Renormalization of the product of two operators $O_1$} \label{Sec:renO1O1}
We now want to apply this procedure in order to derive the renormalization constants
for the insertion of two operators $\widetilde{O}_1$ into a Greens function.
First (using \eqref{D1ZR1}) we notice that
\be\textbf{D}_1\,\textbf{D}_1\,Z^{\ssst{R}}+i\,\textbf{D}_1\,Z^{\ssst{R}}=
\textbf{R}\, \int\!\mathrm{d}\Phi \,\widetilde{O}_1\,\widetilde{O}_1\,\mathrm{e}^{i\int\!\mathrm{d}^4x\,(\ssL+J\cdot\Phi)}.
\ee
On the other hand 
\bea \lb\textbf{D}_1\,\textbf{D}_1+i\textbf{D}_1\rb\,Z^{\ssst{R}} &=&
\lb \textbf{D}_1\,\textbf{D}_1+i\,\textbf{D}_1\rb\,\int\!\mathrm{d}\Phi \,
\mathrm{e}^{i\int\!\mathrm{d}^4x\,(\ssL_{\ssst{B}}+J\cdot\Phi)} \nonumber\\
&=& \lb \textbf{D}_1+i\rb\,\int\!\mathrm{d}\Phi \,
\left\{\sum\limits_{i=1}^{5} Z_{1i}\widetilde{O}_i^{\ssst{B}}\right\}\,
\mathrm{e}^{i\int\!\mathrm{d}^4x\,(\ssL_{\ssst{B}}+J\cdot\Phi)}\nonumber\\
&=& \sum\limits_{i,j=1}^{5} Z_{1i}Z_{1j} 
\int\!\mathrm{d}\Phi \,\widetilde{O}_i^{\ssst{B}}\widetilde{O}_j^{\ssst{B}}\,
\mathrm{e}^{i\int\!\mathrm{d}^4x\,(\ssL_{\ssst{B}}+J\cdot\Phi)} \label{D1D1ZR}\\
&+& \sum\limits_{i=1}^{5} \lb \lb\textbf{D}_1 Z_{1i}\rb + i Z_{1i} \rb\,
\int\!\mathrm{d}\Phi \,\widetilde{O}_i^{\ssst{B}}\,
\mathrm{e}^{i\int\!\mathrm{d}^4x\,(\ssL_{\ssst{B}}+J\cdot\Phi)} \nonumber\\
&+& \sum\limits_{i=1}^{5} Z_{1i} \,
\int\!\mathrm{d}\Phi \,\lb\textbf{D}_1\widetilde{O}_i^{\ssst{B}}\rb\,
\mathrm{e}^{i\int\!\mathrm{d}^4x\,(\ssL_{\ssst{B}}+J\cdot\Phi)}. \nonumber
\eea
This means that appart from the expected term
$\sum\limits_{i,j=1}^{5} Z_{1i}Z_{1j}  O_i^{\ssst{B}} O_j^{\ssst{B}} = [O_1] [O_1]$
linear (L) terms of the form 
$i\,\sum\limits_{i} Z_{1i}^{\ssst{L}}  O_i^{\ssst{B}}$
with new renormalization constants $Z_{1i}^{\ssst{L}}$
will in general contribute to the renormalization of an operator product: 
\be [O_1(x)O_1(0)]=[O_1(x)] [O_1(0)]+i\delta(x) \sum\limits_{i} Z_{1i}^{\ssst{L}}  O_i^{\ssst{B}}(0). \label{RenOP}\ee
A renormalized correlator should hence be defined as
\be i\int\!\mathrm{d}^4x\,e^{iqx} T\{\,[O_1(x)O_1(0)]\}=
i\int\!\mathrm{d}^4x\,e^{iqx} T\{\,[O_1(x)] [O_1(0)] \}-
\sum\limits_{i} Z_{1i}^{\ssst{L}}  O_i^{\ssst{B}}. \label{C1finitecorrO1O1}\ee

We can again compute the first or second line of \eqref{D1D1ZR} and collect 
all fields and renormalization constants into local operators. 
In order to simplify the calculation we note that
\be \int\!\mathrm{d}^4x\,\ssL_{\ssst{B}}= \int\!\mathrm{d}^4x\lb O_1^{\ssst{B}}(x)
+O_3^{\ssst{B}}(x)-O_5^{\ssst{B}}(x)-\f{1}{2\lambda\gs^2}\lb\p_\mu A^{a\,\mu}(x)\rb^2 \rb.
\ee From \eqref{D1toLB} and \eqref{D1toLBZi} we find
\be \textbf{D}_1 \lb \widetilde{O}_1^{\ssst{B}}+\widetilde{O}_3^{\ssst{B}}-\widetilde{O}_5^{\ssst{B}} \rb=
\sum\limits_{i} Z_{1i} \widetilde{O}_i \ee
We solve this for $\textbf{D}_1 \lb \widetilde{O}_1^{\ssst{B}}\rb$ and plug it into the last three lines
of \eqref{D1D1ZR}. Then we discard all unphysical operators as well as their derivatives wrt to
$\gs$ and $\lambda$ as these will not contribute to physical matrix elements.\footnote{For a full set of renormalization constants including the unphysical ones it is necessary to 
extend the set of unphysical operators as not all derivatives of $O_3^{\ssst{B}}$, $O_4^{\ssst{B}}$, $O_5^{\ssst{B}}$
wrt $\gs$ and $\lambda$ can be reabsorbed in exactly these operators.} This yields the result 
\bea
Z_{11}^{\ssst{L}}&=&
\frac{\gs^2}{2} \frac{Z_g''(\gs)}{Z_g}
-\frac{3 \gs^2}{2} \lb \frac{Z_g'(\gs)}{Z_g} \rb^2
-\frac{\gs}{2} \frac{Z_g'(\gs)}{Z_g} \label{Z11L} \\
 &=&-2 \als^2 \lb \f{Z_{\als}'(\als)}{Z_{\als}}\rb^2 + \als^2 \f{Z_{\als}''(\als)}{Z_{\als}}.
\eea

We showed that the idea of \cite{Spiridonov:1984br} for the derivation of renormalization constants
for dimension four operators can also be used for the derivation of renormalization constants
of two such operator insertions. This method can be used for any operator as long as a combination
of derivatives wrt to external currents and parameters of the theory exists which produces
an insertion of this operator into a Greens function starting from the generating functional $Z_R$ of the theory.
In general, care has to be taken that contributions to different physical and unphysical 
operators are separated. Here we considered only one physical operator and discarded the unphysical ones.

Note that the procedure of inserting zero-momentum operators into $Z_R$
will produce only renormalization constants which are momentum independent and do not vanish for $q\to 0$.
Hence we do not find a counterterm here which absorbs the contact terms in $C_0^{\ssst{GG}}$ of 
\eqref{O1O1OPE}. Such a counterterm $Z_0$ is $\propto q^4$ in momentum space. Accounting for this we can
complete \eqref{C1finitecorrO1O1} and \eqref{RenOP}:
\be i\int\!\mathrm{d}^4x\,e^{iqx} T\{\,[O_1(x)O_1(0)]_{\text{full}}\} =
i\int\!\mathrm{d}^4x\,e^{iqx} T\{\,[O_1(x)] [O_1(0)] \}-
\sum\limits_{i} Z_{1i}^{\ssst{L}}  O_i^{\ssst{B}}- q^4 Z_0{}, \label{C0finitecorrO1O1} \ee
\be [O_1(x)O_1(0)] =
 [O_1(x)] [O_1(0)]+i\delta(x) Z_{1i}^{\ssst{L}}  O_i^{\ssst{B}}(0)
 + i \square_{x}^2\delta(x) Z_0{}. \label{fullRenOP}
\ee
But deriving $Z_0$ from first principles is not within the reach this method.

\section{Application to the OPE of the $O_1O_1-$correlator} \label{Sec:ApplO1O1}

In \cite{Zoller:2012qv,Zoller:2014dca} the Wilson coefficient $C_1^{\ssst{GG}}(q^2)$
was computed using the method of projectors \cite{Gorishnii:1983su,Gorishnii:1986gn}.
A projector ${\bf P}$ is applied to both sides of \eqref{O1O1OPE} which has the property
${\bf P}\{O_1^{\ssst{B}}\}=1$ and ${\bf P}\{O_i^{\ssst{B}}\}=0$ for $i\neq 1$.
Thus we computed the bare Wilson coefficient via
\be 
{\bf P}\{i\int\!\mathrm{d}^4x\,e^{iqx} T\{\,[O_1](x)[O_1](0)\}\}=
\sum \limits_i C_{i}^{{\ssst{B}}}(q)\, {\bf P}\{O_{i}^{\ssst{B}}\} \equiv C_{1}^{\ssst{B}}(q).\\
 \label{proj2OO}
\ee
with the projector {\bf P} defined as:\footnote{The Feynman diagram was drawn with the 
Latex package Axodraw \cite{Vermaseren:1994je}.}
\be
C_{1}^{\ssst{B}}(q) =\f{\delta^{ab}}\Ng\f{g^{\mu_1 \mu_2}}{(D-1)}
\f{1}{D}\f{\p}{\p k_1} \cd \f{\p}{\p k_2} \left. \left[
  \begin{picture}(165,50) (0,0)
    \SetWidth{0.5}
    \SetColor{Black}
    \Gluon(10,0)(50,0){5.5}{4.5}
    \Gluon(110,0)(150,0){5.5}{4.5}
    \LongArrow(35,15)(25,15)
    \LongArrow(125,15)(135,15)
\DashCArc(80,0)(56,45,135){4}
\Photon(80,0)(120,40){3}{4}
\Photon(80,0)(40,40){3}{4}
    \CCirc(80,0){30}{Black}{Blue}
    \SetColor{Red}
\Vertex(110,0){4}
\Vertex(50,0){4}
    \SetColor{Black}
    \Text(25,17)[lb]{\Large{\Black{$k_1$}}}
    \Text(125,17)[lb]{\Large{\Black{$k_2$}}}
    \Text(36,-17)[lb]{\Large{\Red{$g_B$}}}
    \Text(110,-17)[lb]{\Large{\Red{$g_B$}}}
    \Text(150,-20)[lb]{\Large{\Black{$\mu_2$}}}
    \Text(5,-20)[lb]{\Large{\Black{$\mu_1$}}}
    \Text(5,10)[lb]{\Large{\Black{$a$}}}
    \Text(150,10)[lb]{\Large{\Black{$b$}}}
  \end{picture}
\right] \right|_{k_i=0}
{}, \label{projectorC1pic}
\ee
where the blue circle represents the the sum of all bare Feynman diagrams
which become 1PI after formal gluing (depicted as a dotted line in \eqref{projectorC1pic})
of the two external lines representing the operators on the lhs of the OPE. 
These external legs carry the large Euclidean momentum q. 
If we use the fully renormalized current \eqref{C1finitecorrO1O1} we find
\be 
{\bf P}\{i\int\!\mathrm{d}^4x\,e^{iqx} T\{\,[O_1(x)O_1](0)\}\}=
\sum \limits_i C_{i}^{{\ssst{B}}}(q)\, 
{\bf P}\{O_{i}^{\ssst{B}}\} -
\sum\limits_{i} Z_{1i}^{\ssst{L}} {\bf P}\{O_{i}^{\ssst{B}}\}
\equiv C_{1}^{\ssst{B}}(q)-Z_{11}^{\ssst{L}}\\
 \label{proj2OOZL}
\ee
and using \eqref{O1ren} we find a fully renormalized Wilson coefficient
as
\be C_1^{\text{ren}}= \f{1}{Z_{11}}C_{1}^{\ssst{B}}(q)-\f{Z_{11}^{\ssst{L}}}{Z_{11}}
\ee

From \eqref{Z11fromZa} and \eqref{Z11L} we compute\footnote{For this we need the QCD $\beta$-function at
three-loop order \cite{Tarasov:1980au,Larin:1993tp}. All given results are in the $\overline{\text{MS}}$-scheme. We define
\mbox{$\as=\f{\als}{\pi}=\f{\gs^2}{4\pi^2}$} and \mbox{$l_{\sss{\mu q}}=\ln\lb\f{\mu^2}{-q^2}\rb$},
where $\mu$ is the $\overline{\text{MS}}$ renormalization scale.
The number of active quark flavours is denoted by $\Nf$,
$\cf$ and $\ca$ are the quadratic Casimir operators of the quark and the adjoint representation of the gauge group
and $\Ng$ is the dimension of the adjoint representation.
$\tr$ is defined through the relation \mbox{$\textbf{Tr}\lb T^a T^b\rb=\tr \delta^{ab}$}.}
\bea 
\f{Z_{11}^{\ssst{L}}}{Z_{11}}&=&
\f{\as^2}{\eps}\left[
      -\frac{17 \ca^2}{24 }
      +\frac{5 \ca   \Nf \tr}{12 }
      +\frac{\cf \Nf \tr}{4 }\right]
     \label{Z11LZ11_3l}\\ &{}&
+\f{\as^3}{\eps}\left[
\frac{1415 \ca^2 \Nf \tr}{864 }
-\frac{2857 \ca^3}{1728 }
+\frac{205 \ca   \cf \Nf \tr}{288 }
-\frac{79 \ca \Nf^2 \tr^2}{432 }
-\frac{\cf^2 \Nf \tr}{16 }
-\frac{11 \cf \Nf^2 \tr^2}{72 }   \right]              \nonumber\\&{}&
+\f{\as^3}{\eps^2}\left[
-\frac{89 \ca^2 \Nf \tr}{144   }
+\frac{187 \ca^3}{288 }
-\frac{11 \ca \cf \Nf \tr}{48 }
+\frac{5 \ca \Nf^2   \tr^2}{36 }
+\frac{\cf \Nf^2   \tr^2}{12 } \right] 
\nonumber
\eea
which is exactly the contact term observed in \cite{Zoller:2012qv,Zoller:2014dca}. Using
\be \als \f{Z_{\als}'(\als)}{Z_{\als}}=-\f{\beta(\als)}{\beta(\als)-\eps} \ee
we arrive at 
\be \f{Z_{11}^{\ssst{L}}}{Z_{11}}=
-\as^2 \f{\beta_1}{\eps}
+\as^3 \lb \f{\beta_0 \beta_1}{\eps^2} - \f{2 \beta_2}{\eps} \rb + 
\as^4 \lb-\f{\beta_0^2 \beta_1}{\eps^3} + \f{\beta_1^2 + 2 \beta_0 \beta_2}{\eps^2}
- \f{3 \beta_3}{\eps} \rb +\mathcal{O}(\als^5)
\ee
and as already suspected in \cite{Zoller:2014dca} the contact term $\f{Z_{11}^{\ssst{L}}}{Z_{11}}$
or $Z_{11}^{\ssst{L}}$ can indeed be expressed through the QCD $\beta$-function to all orders,
namely
\bea
Z_{11}^{\ssst{L}} &=& \f{1}{\eps}\f{-\beta(\als)+ \als\beta'(\als)}{\lb 1-\f{\beta(\als)}{\eps} \rb^2}
=\f{1}{\eps}\lb 1-\f{\beta(\als)}{\eps} \rb^{-2}\als^2\f{\partial}{\partial\als}\left[\f{\beta(\als)}{\als}\right], \label{Z11Lbeta}\\
\f{Z_{11}^{\ssst{L}}}{Z_{11}} &=& \f{1}{\eps}\f{-\beta(\als)+ \als\beta'(\als)}{ 1-\f{\beta(\als)}{\eps} }
=\f{1}{\eps}\lb 1-\f{\beta(\als)}{\eps} \rb^{-1}\als^2\f{\partial}{\partial\als}\left[\f{\beta(\als)}{\als}\right]
.
\eea

Following the prescription of \cite{Chetyrkin:2010dx} we can derive the anomalous dimensions of the Wilson coefficients of the correlator
\eqref{C0finitecorrO1O1} written as\footnote{The renormalization scale $\mu$, which was omitted before for convenience, is carefully reintroduced. $D=4-2\eps$
is the space-time dimension and $O_1^{\ssst{B}}(x)$ has mass dimension $D$.}
\be \widehat{\Pi}^{\ssst{GG}}_{\ssst{full}} :=
i Z_{11}^2\int\!\mathrm{d}^Dx\,e^{iqx} T\{\,O_1^{\ssst{B}}(x) O_1^{\ssst{B}}(0) \}-
 Z_{11}^{\ssst{L}}  O_1^{\ssst{B}}- (\mu)^{-2\eps}q^4 Z_0{}, \label{C0finitecorrO1O1mu} \ee
We find 
\be \mu^2\f{d}{d\mu^2}\widehat{\Pi}^{\ssst{GG}}_{\ssst{full}}=2\gamma_{11}\, \widehat{\Pi}^{\ssst{GG}}_{\ssst{full}}
+\gamma_{11}^L\, [O_1] + \gamma_0\, q^4 \lb \mu^{-2\eps} \rb 
\ee
and the anomalous dimensions are found to be
\bea
\gamma_{11} &=&\mu^2\f{d\log Z_{11}}{d\mu^2} , \label{gamma11}\\
\gamma_{11}^L &=& \lb -\mu^2\f{dZ_{11}^L}{d\mu^2} +2\gamma_{11} Z_{11}^L \rb \f{1}{Z_{11}}\label{gamma11L} ,\\
\gamma_{0} &=& -\mu^2\f{dZ_{0}}{d\mu^2}+(2\gamma_{11}+\eps)Z_0. \label{gamma0}
\eea
Applying these equations and the the well-known relation
\be
\mu^2\f{d}{d\mu^2}=\als\lb\beta(\als)-\eps\rb\f{\partial}{\partial\als} + \mu^2\f{\partial}{\partial\mu^2}
\ee
to \eqref{Z11Lbeta} and \eqref{Z11fromZa} we find (in the limit $\eps\to 0$):
\bea
\gamma_{11} &=& -\als \f{\partial}{\partial\als} \beta(\als), \label{gamma11beta}\\
\gamma_{11}^L &=& \als^2 \f{\partial^2}{\partial\als^2} \beta(\als) \label{gamma11Lbeta},
\eea
where \eqref{gamma11beta} is in agreement with  \cite{Spiridonov:1984br}. For $\gamma_0$ we cannot give a closed 
formula but a three-loop result. In the context of previous calculations \cite{Zoller:2012qv,Zoller:2014dca} we computed the contact term of $C_0^{\ssst{GG}}$, which equals $Z_0$, 
up to three-loop level:\footnote{Only the Adler function of $C_0^{\ssst{GG}}$ with the gauge group factors set to their QCD values was presented explicitly in \cite{Zoller:2012qv}.}
\be \begin{split}
Z_0=  &  \f{\Ng}{16\pi^2}\left\{ \frac{1}{4\eps}
+\frac{\as}{\eps} \left(\frac{17
   \ca }{32}-\frac{5  \Nf
   \tr}{24}\right)    \right.\\ &\left.
+\frac{\as^2}{\eps} \left(\frac{11}{96}
   \ca^2  \zeta_3+\frac{22351 \ca^2
   }{20736}-\frac{7}{24} \ca  \Nf \tr
   \zeta_3-\frac{799 \ca  \Nf
   \tr}{1296} \right.\right. \\ &\left.\left.
   +\frac{1}{4} {\cf}  \Nf \tr
   \zeta_3-\frac{107}{288} {\cf}  \Nf \tr+\frac{49
    \Nf^2 \tr^2}{1296}\right) \right.\\ &\left.
   +\frac{\as}{\eps^2} \left(\frac{ \Nf
   \tr}{12}-\frac{11 \ca
   }{48}\right)  \right.\\ &\left.
   +\frac{{\as}^2}{\eps^2} \left(-\frac{833 \ca^2 }{1152}+\frac{73}{144}
   \ca  \Nf \tr+\frac{1}{12} {\cf} 
   \Nf \tr-\frac{5}{72}  \Nf^2
   \tr^2\right)   \right.\\ &\left.
   +\frac{{\as}^2}{\eps^3}
   \left(\frac{121 \ca^2 }{576}-\frac{11}{72} \ca 
   \Nf \tr+\frac{1}{36}  \Nf^2
   \tr^2\right)\right\}{}, \end{split}
  \ee
  the full Wilson coefficient being 
  \be  \begin{split}
 C_0^{\ssst{GG}} &=  Z_0 + \f{\Ng}{16\pi^2}\left\{ \frac{l_{\sss{\mu q}}}{4}+\frac{1}{4}  \right.\\ &\left.
 +\as \left(\frac{11}{48} \ca   l_{\sss{\mu q}}^2 
 +\frac{73 \ca l_{\sss{\mu q}}   }{48}
   -\frac{3 \ca  \zeta_3}{4}+\frac{485 \ca   }{192} \right. \right. \\  & \left.\left.
   -\frac{1}{12} l_{\sss{\mu q}}^2  \Nf   \tr
   -\frac{7}{12} l_{\sss{\mu q}}  \Nf \tr
   -\frac{17    \Nf \tr}{16}\right)  \right.\\ &\left.
  +\as^2 \left(\frac{121}{576} \ca^2 l_{\sss{\mu q}}^3   
  +\frac{313}{128} \ca^2 l_{\sss{\mu q}}^2   
  -\frac{55}{32} \ca^2 l_{\sss{\mu q}}    \zeta_3
  +\frac{37631 \ca^2 l_{\sss{\mu q}}   }{3456}  \right. \right. \\  & \left.\left.
   -\frac{2059}{288} \ca^2    \zeta_3  
   +\frac{11}{64} \ca^2  \zeta_4
   +\frac{25}{16}   \ca^2  {\zeta5}
   +\frac{707201 \ca^2   }{41472} \right. \right. \\  & \left.\left.
   -\frac{11}{72} \ca l_{\sss{\mu q}}^3    \Nf \tr
   -\frac{85}{48} \ca l_{\sss{\mu q}}^2    \Nf \tr
   -\frac{1}{8} \ca l_{\sss{\mu q}}  \Nf   \tr \zeta_3 \right. \right. \\  & \left.\left.
   -\frac{6665}{864} \ca l_{\sss{\mu q}}    \Nf \tr 
   +\frac{169}{144} \ca  \Nf \tr   \zeta_3
   -\frac{7}{16} \ca  \Nf \tr   \zeta_4 \right. \right. \\  & \left.\left.
   -\frac{7847}{648} \ca  \Nf   \tr
   -\frac{1}{8} {\cf} l_{\sss{\mu q}}^2  \Nf   \tr
   +\frac{3}{4} {\cf} l_{\sss{\mu q}}  \Nf \tr   \zeta_3 \right. \right. \\  & \left.\left.
   -\frac{131}{96} {\cf} l_{\sss{\mu q}}  \Nf   \tr 
   +\frac{41}{24} {\cf}  \Nf \tr   \zeta_3
   +\frac{3}{8} {\cf}  \Nf \tr   \zeta_4 \right. \right. \\  & \left.\left.
   -\frac{5281 {\cf}  \Nf   \tr}{1728}
   +\frac{1}{36} l_{\sss{\mu q}}^3  \Nf^2   \tr^2
   +\frac{7}{24} l_{\sss{\mu q}}^2  \Nf^2   \tr^2\right. \right. \\  & \left.\left.
   +\frac{127}{108} l_{\sss{\mu q}}  \Nf^2   \tr^2 
   +\frac{4715  \Nf^2   \tr^2}{2592}\right) \right\},
 \end{split}
  \ee
  which is known from \cite{Kataev:1982gr},\cite{Chetyrkin:1997iv} and \cite{Baikov:2006ch} for the special case of the gauge group factors replaced by their QCD values.
  This leads to
  \be \begin{split} \gamma_0 &=\frac{\Ng}{4}
  +\frac{1}{48} \as \Ng (51 \ca-20
   \Nf \tr) \\ &
  +\frac{\as^2 \Ng}{6912} \left(\ca^2 (2376 \zeta_3+22351)-16
   \ca \Nf \tr (378 \zeta_3+799) \right. \\ & \left. 
   +8 \Nf \tr (648
   {\cf} \zeta_3-963 {\cf}+98 \Nf
   \tr)\right).
    \end{split}   \ee
\section{Conclusions}

I have presented a derivation of an additive counterterm $Z_{11}^{\ssst{L}} O_1^{\ssst{B}}$
needed to renormalize the correlator of two scalar gluonic operators $O_1$ using the path integral
formalism and extending the ideas of \cite{Spiridonov:1984br}. This counterterm explains
and absorbs the divergences found in the Wilson coefficient $C_1^{\ssst{GG}}$ in 
\cite{Zoller:2012qv,Zoller:2014dca}. A simple closed formula expressing $Z_{11}^{\ssst{L}}$ to
all orders through the QCD $\beta$-function was presented as well. Finally, the anomalous dimensions
$\gamma_{11}$ and $\gamma_{11}^L$ for the correlator of two operators $O_1$ were expressed through the QCD $\beta$-function
to all orders and the anomalous dimension $\gamma_{0}$ was computed at three-loop order.

\section*{Acknowledgments}
I thank K.~G.~Chetyrkin for many useful discussions, for his comments on this paper and his collaboration on 
the previous project \cite{Zoller:2012qv}. 

This research was supported in part by the Swiss National Science Foundation (SNF) 
under contract BSCGI0\_157722.

\bibliographystyle{JHEP}

\bibliography{LiteraturOPETT}

\end{document}